# Riemannian Quantum Circuit

R. V. Ramos and F. V. Mendes

rubens@deti.ufc.br, fernandovm@deti.ufc.br

*Lab. of Quantum Information Technology, Department of Teleinformatic Engineering – Federal University of Ceara - DETI/UFC, C.P. 6007 – Campus do Pici - 60455-970 Fortaleza-Ce, Brazil.*

In this work we present the theory of a unitary matrix related to a finite number of zeros of the Riemann-zeta function. The equivalent quantum circuit and the calculation of the entanglement of a multipartite quantum state produced by the Riemannian quantum circuit are also shown.

## 1. Introduction

Recently, there has been a growing interest in quantum systems related to number theory problems [1-3]. Such interest comes from the early days of quantum mechanics, when Hilbert and Pólya discussed a possible physical solution for Riemann's hypothesis: the zeros of the Riemann-zeta function could be the spectrum of an operator $R = I/2 + iH$, where $H$ is self-ajoint and interpreted as a Hamiltonian. Nowadays, several physical systems related to the zeros of the Riemann-zeta function have been discussed [4,5]. In particular, in [6] the authors, having a finite number of zeros of the Riemann-zeta function, used a numerical method for finding a quantum potential able to reproduce those zeros as energy eigenvalues.

In this work, we show how to construct a quantum circuit, hereafter named Riemannian quantum circuit, whose equivalent unitary matrix has eigenvalues related to the zeros (the amount of zeros considered is equal to the dimension of the unitary matrix) of the Riemann-zeta function. The existence of such quantum circuit implies that, at least in principle, it is always possible to construct a physical system related to any finite amount of zeros using a quantum computer. Additionally, we also show a quantum algorithm based on the Riemannian quantum circuit and we briefly discuss the amount of bipartite entanglement generated by the Riemannian quantum circuit for a particular state having up to 16 qubits.

The present work is outlined as follows: Section 2 brings the procedure for building a unitary matrix whose eigenvalues are related to the zeros of the Riemann-zeta function; Section 3 discusses some applications of the Riemannian quantum circuit; Section 4 shows a quantum circuit related to the unitary matrix obtained in Section 2; at last, conclusions are drawn in Section 5.

## 2. Procedure to construct the Riemannian unitary matrix

Let $s_1$, $s_2$, $s_3$, …, $s_k$, be a set of the first $k$ non-trivial zeros of the Riemann-zeta function, a function of a complex variable $s$ that analytically continues the sum of the infinite serie $\zeta(s) = \sum_n (1/n^s)$. It is always possible to build a ($k \times k$) unitary matrix whose eigenvalues are $s_j^*/s_j$ for $j=1,\ldots,k$. Initially, let us introduce the ($k \times k$) matrix $G$,

$$G = \begin{bmatrix} s_k & 0 & 0 & 0 & 0 & 0 & 0 \\ 0 & s_{k-1} & 0 & 0 & 0 & 0 & 0 \\ 0 & 0 & s_{k-2} & 0 & 0 & 0 & 0 \\ 0 & 0 & 0 & \ddots & 0 & 0 & 0 \\ 0 & 0 & 0 & 0 & s_3 & \vdots & \vdots \\ \vdots & \vdots & \vdots & \vdots & \cdots & (s_1+s_2)/2 & (s_1-s_2)/2 \\ 0 & 0 & 0 & 0 & \cdots & (s_1-s_2)/2 & (s_1+s_2)/2 \end{bmatrix}. \qquad (1)$$

The $k$ eigenvalues of $G$ are the zeros $s_1, s_2, s_3, \ldots, s_k$. Writing the zeros in the form $s_j = a + ib_j$, where both $a$ (as it will be explained latter, the proposed method works only for zeros with the same real part) and $b_j$ are real numbers, the $G$ matrix can be rewritten as the sum of two matrices, $G = aI + iB$,

$$G = aI + iB = a\begin{bmatrix} 1 & 0 & 0 & \cdots & 0 \\ 0 & 1 & 0 & \cdots & 0 \\ \vdots & \vdots & \ddots & \vdots & \vdots \\ 0 & 0 & \cdots & 1 & 0 \\ 0 & 0 & \cdots & 0 & 1 \end{bmatrix} + i\begin{bmatrix} b_k & 0 & 0 & \cdots & 0 \\ 0 & b_{k-1} & 0 & \cdots & 0 \\ \vdots & \vdots & \ddots & \vdots & \vdots \\ 0 & 0 & \cdots & (b_1+b_2)/2 & (b_1-b_2)/2 \\ 0 & 0 & \cdots & (b_1-b_2)/2 & (b_1+b_2)/2 \end{bmatrix}. \qquad (2)$$

The matrix $G' = (1/a)G = I + i(1/a)B$ has eigenvalues $(1/a)s_1$, $(1/a)s_2$, $(1/a)s_3,\ldots,(1/a)s_k$, while the matrix $G'^\dagger = I - i(1/a)B$ has eigenvalues $(1/a)s_1^*, (1/a)s_2^*,\ldots,(1/a)s_k^*$. Since $B/a$ is Hermitean, the Riemannian unitary matrix is simply obtained by

$$U_R = G'^\dagger/G' = \left[I - i(1/a)B\right]/\left[I + i(1/a)B\right]. \qquad (3)$$

The eigenvalues of $U_R$ are exactly $e^{i\theta_j} = s_j^*/s_j$, for $j=1,\ldots,k$. As it can be noted, the procedure just described works only for zeros with the same real part, hence, hereafter we will consider $a = 1/2$. In fact, one may note that, if instead of $G$ given by (1) we had chosen $G$ as a diagonal matrix whose elements are $s_1,\ldots,s_k$, zeros with different real parts (if they exist!) could be used. However, in this case, $G'$ with an identity part could not be obtained and, hence, the unitary matrix $U_R$ could not be constructed. Since $a = 1/2$, one has

$$\theta_j = -\pi + \tan^{-1}\left[-b_j/(1/4 - b_j^2)\right]. \qquad (4)$$

Some information about the angles $\theta_j$ in (4) can be obtained from the Riemann-von Mangoldt formula: the number of Riemann-zeta function zeros $a+ib$ with $0 < b \leq T$ is asymptotically given by $N(T)=(T/2\pi)\log(T/2\pi e)+O(\log(T))$. Now, let us consider the following approximation:

$$\theta_j = -\pi + \tan^{-1}\left[-b_j/\left(1/4 - b_j^2\right)\right] \approx -\pi + \tan^{-1}\left(1/b_j\right) \approx -\pi + 1/b_j. \quad (5)$$

For example, for the lowest zero at the critical line ($b_1 \sim 14.134725142000001$) one has $|\theta_1-(-\pi+1/b_1)| \sim 3 \cdot 10^{-5}$. The amount of $\theta$'s in the range $[\theta_1, \theta_1-\varepsilon]$ (where $\varepsilon << \theta_1$ is a very small angle) is asymptotically given by Riemann-von Mangoldt formula with $T=1/\varepsilon$. The spacing between two consecutive angles $\theta_j$ and $\theta_{j+1}$ is $\Delta\theta_j \sim (1/b_j) - (1/b_{j+1}) = (b_{j+1} - b_j)/(b_{j+1}b_j) \sim (\Delta b_j)(\pi+\theta_{j+1})(\pi+\theta_j)$. However, the spacing between the imaginary part of two consecutive zeros, $\Delta b_j$, is asymptotically given by $2\pi/\log(j)$, hence, $\Delta\theta_j \sim (2\pi/\log(j))(\pi+\theta_{j+1})(\pi+\theta_j)$. A plot of $\Delta\theta$ versus $\theta$ can be seen in Fig. 1. The curve I is obtained using zeros with $b \in [101.3178510060000, 120000.3764067760]$ while the curve II is the analytical formula $(2\pi/\log(j))(\pi+\theta_{j+1})(\pi+\theta_j)$ using $j \in [30,169165]$ (the 30[th] zero is $\frac{1}{2}+i101.3178510060000$ and the 169165[th] zeros is $\frac{1}{2}+i120000.3764067760$).

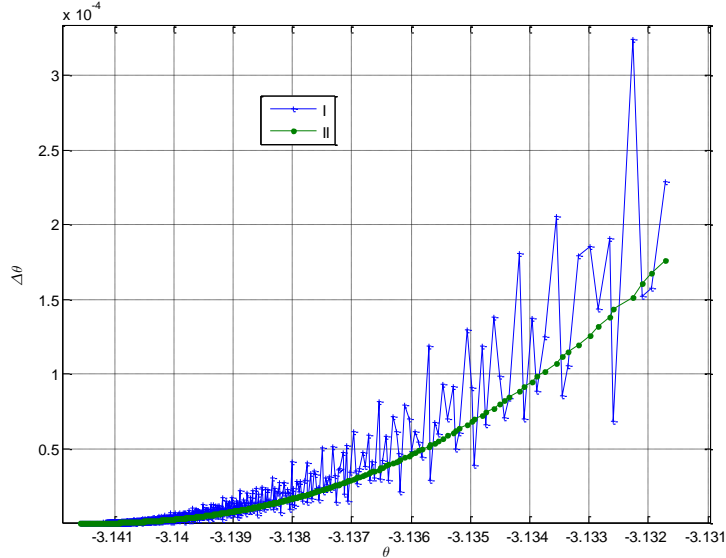

Fig. 1 – Spacing between two consecutive $\theta$'s versus $\theta$: I) Obtained using zeros with $b \in [101.3178510060000, 120000.3764067760]$. II) $(2\pi/\log(j))(\pi+\theta_{j+1})(\pi+\theta_j)$ for $j \in [30,169165]$.

## 3. Applications of the Riemannian quantum circuit

Since the term $-\pi$ in (4) will appear as a global phase in the quantum states,

hereafter it will not be taken into account anymore. Thus, one has that $\theta_j \sim 1/b_j$. Now, let us consider $|\psi_1\rangle,\ldots,|\psi_k\rangle$ the set of eigenvectors of $U_R$ ($k \times k$), where $|\psi_j\rangle$ is the eigenvector associated to the eigenvalue $\exp(i\theta_j)$. Now, let us describe a quantum algorithm for finding the value of $\sum_{j=1}^{k} \theta_j$, where $k$ is the number of zeros. Firstly, we define the quantum translation operator $T$ that shifts the eigenstates by one unity, that is,

$$T|\psi_j\rangle = |\psi_{j+1}\rangle \quad \forall\ 1 \leq j < k \tag{6}$$

$$T|\psi_k\rangle = |\psi_1\rangle. \tag{7}$$

Furthermore, we define the Riemannian state as

$$|\psi_R\rangle = U_R \sum_{j=1}^{k} \frac{1}{\sqrt{k}} |\psi_j\rangle = \sum_{j=1}^{k} \frac{1}{\sqrt{k}} e^{i\theta_j} |\psi_j\rangle. \tag{8}$$

Using (6)-(8) it is straightforward to check that

$$(TU_R)^k \sum_{j=1}^{k} \frac{1}{\sqrt{k}} |\psi_j\rangle = e^{i\left(\sum_{m=1}^{k} \theta_m\right)} \sum_{j=1}^{k} \frac{1}{\sqrt{k}} |\psi_j\rangle. \tag{9}$$

Hence, the quantum state $\left(1/\sqrt{k}\right)\sum_{j=1}^{k}|\psi_j\rangle$ is an eigenvector of the operator $(TU_R)^k$ with eigenvalue $e^{i(\sum_{m=1}^{k}\theta_m)}$. Thus, the quantum eigenvalue estimation algorithm [7] can be used to get an estimation of $\sum_{m=1}^{k}\theta_m$. Similarly, one can develop a quantum algorithm for estimating the value of $\sum_{m=1}^{k-1}\Delta_{m+1,m}$, where $\Delta_{m+1,m}$ is the difference of the angles of two consecutive eigenvalues: $\theta_{m+1} - \theta_m$ (and, hence, $\sum_{m=1}^{k-1}\Delta_{m+1,m} = \theta_k - \theta_1$ ). In this case, one should firstly note that $R|\psi_j\rangle = \exp(\theta_{j+1}-\theta_j)|\psi_j\rangle$, where $R = (TU_R)^{\dagger}(U_RT)$. Thus, $T(TR)^{k-1}|\psi_1\rangle = e^{i(\sum_{m=1}^{k}\Delta_{m+1,m})}|\psi_1\rangle$ and, hence, $\sum_{m=1}^{k-1}\Delta_{m+1,m}$ can be estimated by using the quantum eigenvalue estimation algorithm.

At last, let us to check the multipartite entanglement of the Riemannian state given in Eq. (8), whose representation in the canonical basis (the basis whose vectors are the columns of the identity matrix:$\{|00\ldots00\rangle, |00\ldots01\rangle, |00\ldots10\rangle, \ldots,|11\ldots10\rangle, |11\ldots11\rangle\}$) is,

$$|\psi_R\rangle = \frac{1}{\sqrt{k}} \left[ \begin{array}{c} e^{i\theta_k}|00\ldots00\rangle + e^{i\theta_{k-1}}|00\ldots01\rangle + \ldots + \sqrt{2}e^{i\left(\frac{\theta_2+\theta_1}{2}\right)}\cos\left(\frac{\theta_2-\theta_1}{2}\right)|11\ldots10\rangle \\ + i\sqrt{2}e^{i\left(\frac{\theta_2+\theta_1}{2}\right)}\sin\left(\frac{\theta_2-\theta_1}{2}\right)|11\ldots11\rangle \end{array} \right]. \tag{10}$$

In (10) one has $k = 2^n$, where $n$ is the number of qubits. The Riemannian state $|\psi_R\rangle$ carries information about the first $k$ zeros of the Riemann-zeta function. In order to measure its multipartite entanglement, we are going to use the average bipartite entanglement between all bipartitions of the Riemannian state, $E_1$, and the average bipartite entanglement considering only those bipartitions where one of the parts has only one qubit, $E_2$:

$$E_1 = \frac{1}{2^{n-1}-1} \sum_{k=1}^{2^{n-1}-1} E\left(\rho_{\Omega_k\_\Omega-\Omega_k}\right) \qquad (11)$$

$$E_2 = \frac{1}{n} \sum_{k=1}^{n} E\left(\rho_{\omega_k\_\Omega-\omega_k}\right) \qquad (12)$$

In (11)-(12), $\Omega$ represents the full set of $n$ qubits, $\Omega_k$ represents a particular subset and $\Omega-\Omega_k$ is the subset of $\Omega$ whose elements do not belong to $\Omega_k$. Moreover, $\omega_k$ represents a set of only one qubit and $\Omega-\omega_k$, with $n$-1 elements, is the subset of $\Omega$ whose elements do not belong to $\omega_k$. For instance, for a Riemann state with four qubits one has $\Omega=\{A,B,C,D\}$ and the average entanglements are:

$$E_1 = \frac{1}{7}\left[\begin{array}{l} E\left(\rho_{A\_BCD}\right)+E\left(\rho_{B\_ACD}\right)+E\left(\rho_{C\_ABD}\right)+E\left(\rho_{D\_ABC}\right)+E\left(\rho_{AB\_CD}\right)+ \\ E\left(\rho_{AC\_BD}\right)+E\left(\rho_{AD\_BC}\right) \end{array}\right] \qquad (13)$$

$$E_2 = \frac{1}{4}\left[E\left(\rho_{A\_BCD}\right)+E\left(\rho_{B\_ACD}\right)+E\left(\rho_{C\_ABD}\right)+E\left(\rho_{D\_ABC}\right)\right] \qquad (14)$$

The entanglements of $|\psi_R\rangle$ versus number of qubits (up to 16 qubits) are shown in Fig. 2. There, $Ent_1$ and $Ent_3$ are obtained from (13) and (14), respectively, by using the von Neumann entropy as bipartite entanglement measure: $E(\rho_{x\_y}) = S_{VN}(\rho_x) = -Tr[\rho_x \log(\rho_x)]$, where $\rho_x = Tr_y(\rho_{x\_y})$ and $Tr$ means the partial trace operation. Similarly, $Ent_2$ and $Ent_4$ are obtained from (13) and (14), respectively, by using the linear entropy as bipartite entanglement measure: $E(\rho_{x\_y}) = S_L(\rho_x) = 2\{1-Tr[(\rho_x)^2]\}$. For example, considering the von Neumann entropy, for the four qubit case one has $E(\rho_{A\_BCD}) = S_{VN}(\rho_A) = S_{VN}(Tr_{BCD}(|\psi_R\rangle\langle_R\psi|))$ and $E(\rho_{AB\_CD}) = S_{VN}(\rho_{AB}) = S_{VN}(Tr_{CD}(|\psi_R\rangle\langle_R\psi|))$.

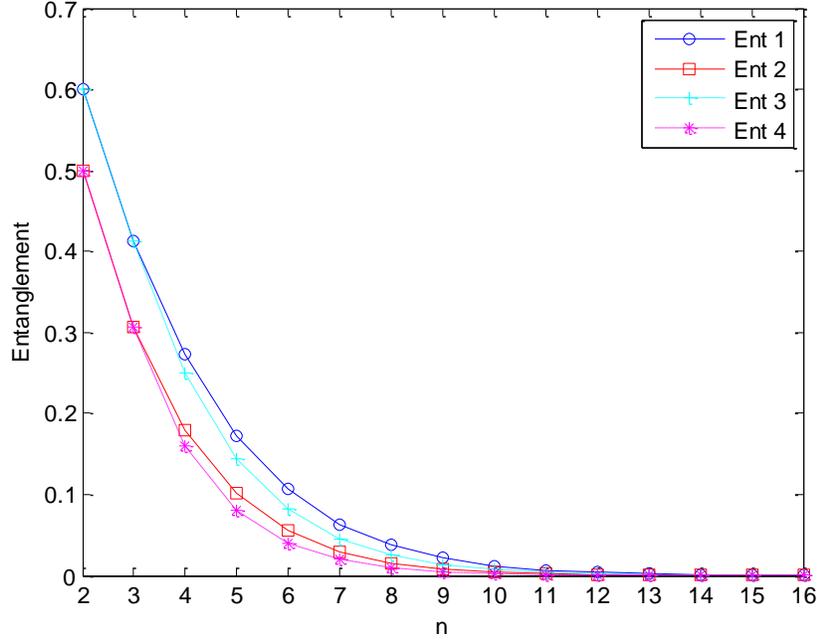

Fig. 2 – Average bipartite entanglements versus number of qubits for the Riemannian state given in (10).

As it can be noted, the larger the number of qubits ($n$) the lower is the average entanglement. In order to understand this behavior, let us consider the distance between the Riemann state and the disentangled state $H^{\otimes n}|0\rangle^{\otimes n} = 1/\sqrt{2^n}\sum_{x=00...0}^{11...1}|x\rangle$: the smaller the distance the lower is the entanglement. Since both of them are pure states, the distance can be simply given by the fidelity: $F = |\langle\psi_R|H^{\otimes n}|0\rangle^{\otimes n}|^2$, with $k = 2^n$. We are going to show that $F$ tends to the value 1 (meaning that they are indistinguishable) when $n$ grows. Firstly, one can note that

$$\left|\sum_{j=0}^{k}e^{i\theta_j}\right|^2 = \sum_{j,l=1}^{k}e^{i(\theta_j-\theta_l)} = \sum_{j=1}^{k}1 + 2\sum_{\substack{j,l=1\\j>l}}^{k}\cos(\theta_j-\theta_l) \approx k + 2\sum_{\substack{j,l=0\\j>l}}^{k}\cos\left(\frac{1}{b_j}-\frac{1}{b_l}\right)$$

$$\approx k + 2\sum_{\substack{j,l=0\\j>l}}^{k}\left(1-\frac{\left(b_j^{-1}-b_l^{-1}\right)^2}{2}\right) \approx k + 2\sum_{\substack{j,l=0\\j>l}}^{k}1 = k + 2\frac{k(k-1)}{2} = k^2 \qquad (15)$$

In (15) we used $\cos(\phi) \sim 1-\phi^2/2 \sim 1$, since $\phi = \left(b_j^{-1} - b_l^{-1}\right)$ is very small. Now, returning to the fidelity, after some algebra one has that

$$F = \left|\langle 0^{\otimes n} | H^{\otimes n} | \psi_R \rangle\right|^2 = \frac{1}{k^2} \left| \sum_{j=1}^{k} e^{i\theta_j} - e^{i\theta_2} + \left(\sqrt{2}-1\right) e^{i\theta_1} \right|^2. \qquad (16)$$

When *k* grows, the term out of the summation remains constant while the summation grows and become dominant. Hence, using (15) one has that $F \sim 1$ when *k* is large enough, what means that, for a large number of qubits, $|\psi_R\rangle$ is arbitrarily close to a disentangled state, hence, it must have low entanglement. This behavior can be seen in Fig. 3 that shows $|\langle \psi_R | H^{\otimes n} | 0 \rangle^{\otimes n}|^2$ versus number of qubits (up to 16 qubits).

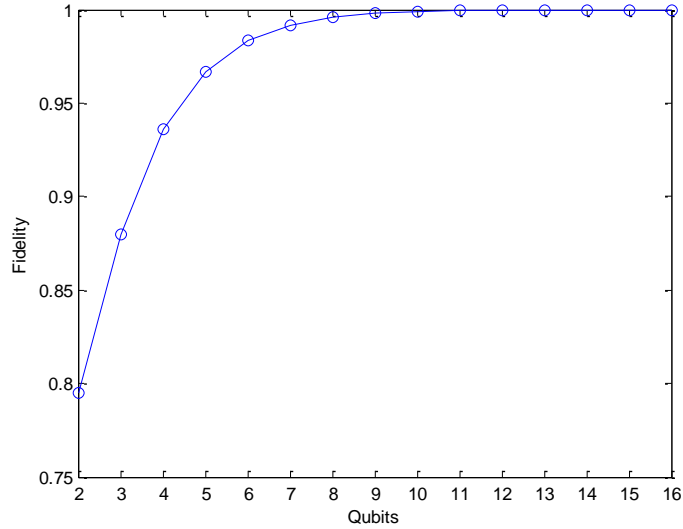

Fig. 3 – Fidelity versus number of qubits. $F = |\langle\psi_R|H^{\otimes n}|0\rangle^{\otimes n}|^2$.

## 4. Procedure to build the Riemannian quantum circuit

Having the (*k* x *k*) unitary matrix given in (3), it is always possible to obtain a quantum circuit that represents its physical realization. For simplicity, here we consider the number of zeros $k = 2^N$ in order to have a quantum circuit for *N* qubits.

One can use different procedures in order to obtain a quantum circuit from a unitary matrix, for example the sin-cos decomposition [8]. In general, different procedures will result in different quantum circuits. For the Riemannian quantum circuit, one is expected to obtain a quantum circuit with several CNOTs and single-qubit gates whose parameters' values depend on the values of the zeros of Riemann-zeta function. Here we adopt a simpler approach, based on sin-cos decomposition, using the eigenvectors of $U_R$: the *N* qubits of the eigenvectors act as controllers in a *N*-qubit controlled gate. Using this, the quantum circuit for $U_R$ is as shown in Fig. 4.

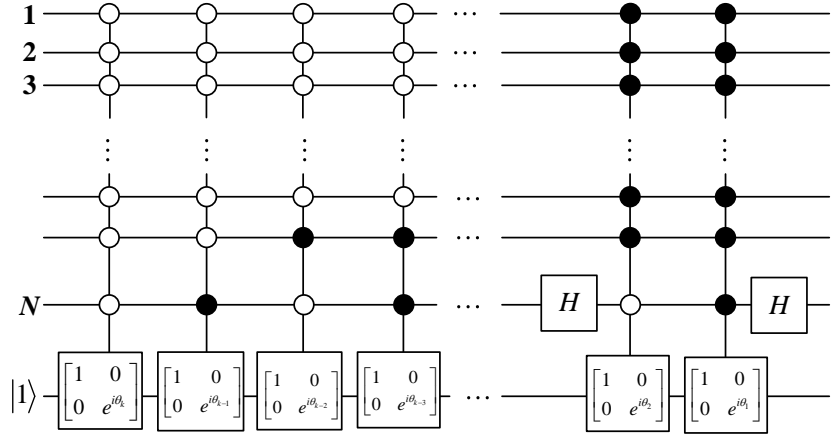

Fig. 4 – Riemannian quantum circuit.

In Fig. 4, $H$ is the Hadamard gate:

$$H = \frac{1}{\sqrt{2}} \begin{bmatrix} 1 & 1 \\ 1 & -1 \end{bmatrix}. \tag{17}$$

The eigenstates of the quantum circuit shown in Fig. 4 are the eigenvectors of $U_R$ (the last qubit is always in the state $|1\rangle$ and it is not considered): $|\psi_k\rangle = |00\ldots000\rangle$, $|\psi_{k-1}\rangle = |00\ldots001\rangle$, $|\psi_{k-2}\rangle = |00\ldots010\rangle$, $|\psi_{k-3}\rangle = |00\ldots011\rangle$,…, $|\psi_2\rangle = |11\ldots11\rangle(|0\rangle+|1\rangle)/2^{1/2}$, $|\psi_1\rangle = |11\ldots11\rangle(|0\rangle-|1\rangle)/2^{1/2}$.

The quantum circuit in Fig. 4 shows how to program (hence it requires the knowledge of the used zeros) a universal quantum computer for working as a physical system related to a finite set of zeros of the Riemann-zeta function.

## 5. Conclusions

Firstly, one can note that our approach is different from the traditional approach found in the literature where people look for a quantum system related to all infinite zeros. In general, the quantum potentials used in such quantum systems are hard to find in nature or to construct artificially. Secondly, we would like to stress that, although we had always considered the first $k$ zeros of the Riemann-zeta function, the theory here described works for any set of $k$ zeros. Hence, we can take any finite number of zeros, anywhere in the critical line, and build a quantum circuit whose eigenvalues are related to them in a very clear way: each eigenvalue depends on only one zero. Hence, one can say that all zeros of the Riemann-zeta function are related to a physical system. In other words, our approach shows how to construct a physical system, with finite resources (finite number of quantum gates), able to work with any (finite) set of different zeros. Furthermore, since this physical system is a quantum circuit, it can be (at least in principle) programmed in a universal quantum computer and physically implemented with optics, superconductor, quantum dots or any other technology for quantum computer implementation.

It can be argued that it is easy to produce a unitary matrix and, consequently, a quantum circuit whose eigenvalues are related to the zeros of the Riemann-zeta function. For example, a ($k$ x $k$) unitary matrix related to the zeros could be a diagonal matrix whose elements are $e^{i\theta_j}$ $j = 1,..,k$, where $\theta_j$ is as given in (4). In this case, the eigenstates are the states of the canonical basis. However, we consider this is not a natural path since the real part of the zeros is not taken into account in any moment. Furthermore, it does not follow the Hilbert-Pólya suggestion of searching for an operator of the type $I/2+iH$. In our approach, the term $I/2$ comes naturally from the zeros.

## Acknowledgments

This work was supported by the Brazilian agency CNPq via Grant no. 303514/2008-6. Also, this work was performed as part of the Brazilian National Institute of Science and Technology for Quantum Information.